\documentstyle[12pt]{article}

\input epsf

\def\bild#1#2{    
        \vspace*{-5mm}
        \begin{center}
        \begin{math}
        \epsfxsize#2cm
        \epsffile{#1}
        \end{math}
        \end{center}
        }

\newcommand{\ppp}[1]{%
        \setbox0=\hbox{#1}%
        \kern-.02em\copy0\kern-\wd0
        \kern+.04em\copy0\kern-\wd0
        \kern-.02em\raise.0217em\box0}
\newcommand{\vek}[1]{
        \mathchoice{\mbox{\boldmath$#1$}}%
        {\mbox{\boldmath$#1$}}%
        {\ppp{$\scriptstyle#1$}}%
        {\ppp{$\scriptscriptstyle#1$}}}

\newcommand{\lsim}{
 \mathrel{\setbox0=\hbox{$<$}\raise0.6ex\copy0\kern-\wd0
 \lower0.65ex\hbox{$\sim$}}}

\newcommand{\gsim}{
 \mathrel{\setbox0=\hbox{$>$}\raise0.6ex\copy0\kern-\wd0
 \lower0.65ex\hbox{$\sim$}}}

\begin{document}
%
\begin{titlepage}
\renewcommand{\thefootnote}{\fnsymbol{footnote}}
\makebox[2cm]{}\\[-1in]
\begin{flushright}
\begin{tabular}{l}
TUM/T39-96-26
\end{tabular}
\end{flushright}
\vskip0.4cm
\begin{center}
  {\Large\bf
    Gluon Polarization from QCD Sum Rules
    \footnote{Work supported in
    part by BMBF}
}\\ 

\vspace{2cm}

L.\ Mankiewicz,\footnote{On leave of absence from N. Copernicus
Astronomical Center, Polish Academy of Science, ul. Bartycka 18,
PL--00-716 Warsaw (Poland)} 
G. Piller and A. Saalfeld

\vspace{1.5cm}

{\em Institut f\"ur Theoretische Physik, TU M\"unchen, Germany}

\vspace{1cm}

{\em \today}

\vspace{1cm}

{\bf Abstract:\\[5pt]} \parbox[t]{\textwidth}{ The gluon polarization $\Delta
  G$ in a nucleon can be defined in a gauge invariant way as the integral over
  the Ioffe-time distribution of polarized gluons. We argue that for
  sufficiently regular polarized gluon distributions $\Delta G$ is dominated by
  contributions from small and moderate values of the Ioffe-time $z\lsim 10$.
  As a consequence $\Delta G$ can be estimated with $20\%$
  accuracy from the first two even moments of the polarized gluon distribution,
  and its behavior at small values of Bjorken $x$ or, equivalently, at large
  Ioffe-times $z$.  We employ this idea and compute the first two moments of
  the polarized gluon distribution within the framework of QCD sum rules.
  Combined with the color coherence hypothesis we obtain an upper limit for
  $\Delta G \sim 2 \pm 0.5$ at a typical scale $\mu^2 \sim 1$ GeV$^2$.  }

\end{center}
\end{titlepage}
\setcounter{footnote}{0}

\newpage

During the last two decades deep-inelastic lepton-nucleon scattering (DIS) has
proven to be one of the most valuable sources of information about nucleon
structure. According to the Operator-Product-Expansion (OPE), the
$Q^2$-dependence of the cross-section can be entirely included in
perturbatively calculable Wilson coefficients, and thus completely factorized
from the effects of long-distance dynamics, described by a set of twist-2
parton distribution functions which are defined at a certain reference scale.
The analysis of experiments with unpolarized lepton beams has provided detailed
informations on quark momentum distribution functions and has lead to the
conclusion that -- even at relatively low scales of few GeV$^2$ -- the nucleon
momentum is shared almost equally between quarks and gluons.  With the
advent of high-quality polarized targets and beams it has become possible to
study polarized quark distributions \cite{Exp}.  In particular, combining
experimental results with SU(3)-flavor symmetry allows to determine the
fraction $\Sigma$ of the nucleon polarization carried by quarks \cite{Ellis96}.
Since the latter turns out to be much smaller, $\Sigma \approx 0.3$, than
expected from simple quark models \cite{Rev}, it is natural to search for other
contributions to the nucleon spin.

At a given scale $\mu^2$ the nucleon spin sum rule allows to 
separate the nucleon spin into gauge-invariant quark and gluon 
contributions \cite{Ji96}:
\begin{equation}
\frac{1}{2} = J_q(\mu^2) + J_g(\mu^2)\, ,
\label{spinSR}
\end{equation}
with both parts calculable as matrix elements of local operators 
sandwiched between polarized nucleon states.
The quark contribution can be split further into gauge-invariant 
polarization and angular momentum parts:
\begin{equation}
J_q(\mu^2)  =  \frac{1}{2} \Sigma(\mu^2) + L_q(\mu^2).
\end{equation}
Here the quark polarization $\Sigma(\mu^2)$ is linked to the matrix 
element  of the flavour singlet axial current operator 
$\int d^3 x \,{\bar\psi} {\vek \gamma} \gamma_5 \psi$, 
and the angular momentum $L_q(\mu^2)$ is connected to the 
gauge-invariant angular momentum operator 
$ - \int d^3 x \,{\bar \psi} (\vek x \times i\vek D) \psi$, 
with the covariant derivative 
$D^{\mu}  = (D_0,\vek D) = \partial^{\mu} + i g A^{\mu}$. 
The separation of the gluon contribution $J_g$ into angular
momentum and polarization parts,  
\begin{equation}
J_g(\mu^2)  =  \Delta G(\mu^2) + L_g(\mu^2), 
\end{equation}
is more involved. The reason is that in QCD appropriate relations
between (even) moments of the polarized gluon density and matrix elements of
twist-2 operators exist  only for $l \ge 2$:
\begin{eqnarray}
\int_0^1 \, dx \, x^l \, \Delta G(x,\mu^2) & \equiv & \Gamma_l(\mu^2),
\quad \mbox{with} \,~l = 2,4,\dots,   
\nonumber \\ 
\langle p,s| n_\mu n_\nu {\mbox Tr}\, G^{\mu}_{~\xi} (in\cdot D)^{l-1} 
{\tilde G}^{\xi\nu} |p,s\rangle  
& = & (s\cdot n) (p\cdot n)^{l}
\Gamma_l(\mu^2).
\label{twist2} 
\end{eqnarray}
Here $x$ stands for the Bjorken scaling variable which is equal to the gluon
light-cone momentum fraction and $G^{\mu\nu}$ is the gluon field strength
tensor.  The nucleon state $|p,s\rangle$ carries an invariant mass $m_N$,
momentum $p$ and spin $s$, with $s^2 = - m_N^2$. The light-like vector $n_\mu$,
$n^2=0$, projects out the ``plus'' component of any four vector, i.e. $n \cdot
a = a^+ \equiv a^0 + a^3$.  There is no corresponding relation for $l=0$ or
$\Delta G(\mu^2) = \int_0^1\, dx\, \Delta G(x,\mu^2)$, because a suitable gauge
invariant, charge conjugation even, local operator does not exist.  Thus
contrary to the quark case, the gluon polarization cannot be extracted from a
matrix element of a local QCD operator.

On the other hand experiments, like large $p_T$ jets or open charm production
in polarized DIS, can in principle provide experimental informations about the
polarized gluon distribution $\Delta G(x,\mu^2)$ (see e.g. \cite{Maul96}) and its
integrated strength $\Delta G$ which, as a measurable quantity, should have a
gauge-invariant definition. Indeed, consider the non-local twist-2 operator
\cite{Man90} 
\begin{equation}
O(\Delta/2;-\Delta/2) = n_\mu n_\nu
{\mbox Tr}\, G^{\mu\xi}(\Delta/2)[\Delta/2;-\Delta/2]
{\tilde G}_\xi^\nu(-\Delta/2),  
\label{defO}
\end{equation}
where $\Delta$ is a light-like vector proportional to $n$.  To ensure gauge
invariance the operator (\ref{defO}) contains the path-ordered exponential in
the adjoint representation $[\Delta/2;-\Delta/2]$.  The nucleon matrix element
of the operator (\ref{defO}), renormalized at a scale $\mu^2$, can be written
in terms of the polarized gluon density $\Delta G(x,\mu^2)$:
\begin{equation} \label{op_io}
\langle p,s | O(\Delta,0) | p,s \rangle = (p\cdot n) (s \cdot n) 
\int_0^1 \, dx \, x \, \Delta G(x,\mu^2) 
\sin\left[x (p\cdot \Delta)\right] \, ,
\end{equation}
which defines the corresponding distribution $\Gamma(z,\mu^2)$ in terms of the
''Ioffe-time'' $z = p \cdot \Delta$ \cite{Iofdst} as:
\begin{equation} \label{DelG-Ioffe}
\Gamma(z,\mu^2) = \int_0^1\, dx\, x\, \Delta G(x,\mu^2) \sin(x z).
\end{equation}
Performing a Taylor expansion of eq.(\ref{op_io}) around $\Delta = 0$ one can
easily see that the definition of the Ioffe-time distribution is equivalent 
to the OPE relation between matrix elements of local operators 
(\ref{twist2}) and moments of $\Delta G(x,\mu^2)$.  Now, although the first 
moment of $\Delta G(x,\mu^2)$ is not determined by a local operator, it can 
be extracted under the assumption that $\Delta G(x,\mu^2)$ is sufficiently 
regular at $x=0$. Using the relation:
\begin{equation}
\int_0^\infty dz \sin (x z) = P \frac{1}{x} \,,
\end{equation}
we obtain:
\begin{equation}
\Delta G(\mu^2) \equiv \int_0^1 dx \,\Delta G(x,\mu^2) = 
\int_0^\infty dz \, \Gamma(z,\mu^2) \, .
\label{defDG}
\end{equation}

In the the Fock-Schwinger gauge the integral (\ref{defDG}) can be related to
the matrix element of the topological current $K^\mu = \frac{\alpha_S}{2 \pi}
\epsilon^{\mu\nu\rho\sigma} A^a_\nu (\partial_\rho A^a _\sigma + \frac{1}{3} g
f_{abc} A^b_\rho A^c_\sigma)$ \cite{BB91}.  Recently  Jaffe \cite{Jaffe96} used
this relation to compute $\Delta G(\mu_B^2)$ in the bag model, taking into
account gluonic contributions arising from one-gluon exchange between quarks.
In a semi-classical framework, i.e.  neglecting renormalization, one obtains a
negative value of $\Delta G \sim - 0.2$ at the bag scale $\mu^2_B$.  In the
present paper we propose another procedure which allows, under favorable
circumstances, to estimate $\Delta G(\mu^2)$ with an accuracy sufficient from
any phenomenological point of view.

Equation (\ref{defDG}) defines the gluon polarization in terms of all higher
moments of the polarized gluon distribution, and as such has a limited
practical value\footnote{Note that the QCD interpretation of (\ref{defDG}) is
  similar to the Gottfried sum rule which also cannot be connected to a matrix
  element of a twist-2 local operator, but is determined by an infinite number
  of matrix elements of a tower of twist-2 operators.}.  We argue, however,
that a reasonable estimate for $\Delta G (\mu^2)$ can be obtained from the
first few moments only, assuming a sufficiently smooth polarized gluon
distribution.  In particular our previous studies \cite{Weigl96} of unpolarized
Ioffe-time distributions have shown that an overall good approximation can be
obtained by combining small and large $z$ expansions and the behavior of parton
distributions at large values of the Bjorken variable $x \to 1$.  At small
longitudinal distances, i.e. small $z$, the Ioffe-time distribution can be
approximated by its Taylor expansion around the origin, with coefficients
calculable as matrix elements of local twist-2 operators of the lowest
dimension.  At large $z$ its shape is sensitive to the physics of large
longitudinal distances, which corresponds to small $x$.  In this region a
proper description of parton distributions in terms of QCD is still under
debate \cite{Lip96}.  An analysis of unpolarized distributions \cite{Weigl96}
shows that the transition between the small and large $z$ domain is relatively
sharp, and that the large longitudinal distance regime prevails beyond $z \sim
10$.  Note that within a classical picture of a nucleon localized at the origin
a Ioffe-time of $z \approx 10$ corresponds to a longitudinal distance of
approximately $2$ fm which, perhaps not surprisingly, coincides with the
classical nucleon size\footnote{Such a picture makes sense only as far as the
  nucleon Compton wavelength is about $5$ times shorter than its
  electromagnetic radius.}.

In Figure 1 and 2 we show Ioffe-time distributions resulting from two very
different polarized gluon densities.  While the distribution of Brodsky et al.
\cite{Bro95} gives $\Delta G = 0.54$ at a typical hadronic scale, the
distribution of Chiappetta et al. \cite{Chiap} leads to $\Delta G = 1.7$ at
$\mu^2 = 1\,GeV^2$.  In both cases the correlation function $\Gamma(z,\mu^2)$
falls off quickly at large $z$ and, due to the resulting triangular shape of
$\Gamma(z,\mu^2)$, the main contribution to the first moment (\ref{defDG}) is
contained in the region $z \lsim z_0$, with $z_0$ around $10$. 
This agrees with the intuition that degrees of freedom
which determine the large $z$ behavior of parton distributions do not couple
to the target polarization. 
Note that if $\Delta G(x,\mu^2)$ behaves like 
$\sim 1/x^\alpha$ one finds $\Gamma(z,\mu^2)
\sim 1/z^{2-\alpha}$.
The dominance of contributions from moderate $z$ 
can be quantified by calculating
the fraction of $\Delta G(\mu^2)$ which results from the interval between $0$
and a given $z$, as shown in Figure 1 and 2.  Integrating up to $z_0 = 10$ one
can determine $\Delta G(\mu^2)$ with 20\% accuracy for the distribution of
Brodsky et al.\cite{Bro95}, while for the distribution of Chiappetta et al.
\cite{Chiap} the accuracy is even better, of the order of 10\%. Note that the
distribution of Ref.\cite{Chiap} yields a Ioffe-time distribution $\Gamma(z)$
which oscillates around zero at large $z$.  Therefore a fraction of the full
integral (\ref{defDG}) obtained by an integration up to a finite value of $z$
can be larger then one. The slightly worse accuracy obtained for the
distribution of Ref.\cite{Bro95} is due to the fact that in this case the
polarization $\Delta G(\mu^2)$ is much smaller and the corresponding Ioffe-time
distribution has a rather flat shape, which is seen when both distributions are
compared at the same vertical scale.

We show now that only a few moments of $\Delta G(x,\mu^2)$ are necessary to
obtain a reasonable estimate of $\Delta G(\mu^2)$. Already the first two
non-vanishing moments are sufficient to obtain the approximation to the
Ioffe-time distribution shown in Figure 1 and 2:
\begin{equation} \label{eq:Ioffe_approx}
\Gamma(z,\mu^2) \approx \Gamma_2(\mu^2) z - \frac{1}{6} \Gamma_4(\mu^2) z^3.
\end{equation}
As an approximation to $\Delta G(\mu^2)$ we calculate the area bound by the
approximate Ioffe-time distribution (\ref{eq:Ioffe_approx}) up to its maximum,
and the straight line connecting this point with the value of
$\Gamma(z=10,\mu^2)$. For the distributions of Ref. \cite{Bro95} and
\cite{Chiap} this procedure yields $\Delta G = 0.41$ and $\Delta G = 1.67$, as
compared to the exact values $0.54$ and $1.7$, respectively.  Although due to
the reasons discussed above the accuracy is lower in the first case, in general
both estimates are quite satisfactory.  In an even simpler
approach, calculating 
the area of the triangle spanned by the points $z=0$, $z=10$, and the
maximum of the approximate 
Ioffe-time distribution (\ref{eq:Ioffe_approx}), one obtains:
\begin{equation}
\Delta G(\mu^2)  \approx \frac{10}{3} \Gamma_2(\mu^2) 
\sqrt{\frac{2 \Gamma_2(\mu^2)}{\Gamma_4(\mu^2)}}\,,
\end{equation}
which yields $\Delta G = 0.28$ for the distribution of Ref.\cite{Bro95} while
for Ref.\cite{Chiap} one obtains $\Delta G =1.1$, i.e. an accuracy of about
50\% in both cases. This demonstrates that the gluon polarization is indeed
dominated by the two first moments of the polarized gluon distribution.  We
note that the whole procedure assumes implicitly that the Ioffe-time polarized
gluon distribution has a regular shape which in principle cannot be taken for
granted.  On the other hand all presently advocated distributions \cite{Lad96}
have this property, and therefore it can be considered at least as a reasonable
hypothesis.

As explained above, besides (at least) the first two moments, we need to know
the normalization of $\Gamma(z,\mu^2)$ at large $z \approx 10$. As the large
$z$ region corresponds to the small $x$ domain, we need information about the
shape and normalization of the small $x$ polarized gluon distribution. There,
the color coherence of gluon couplings results in a simple relation between the
unpolarized and polarized gluon distribution \cite{Bro95}:
\begin{equation}
\left( \frac{\Delta G(x)}{G(x)}\right)_{\mbox{nucleon}}
\rightarrow {x}, 
\quad \mbox{for} \;x\to 0.
\label{coh}
\end{equation}
Using GRV \cite{GRV} and CTEQ \cite{CTEQ} LO unpolarized gluon distributions we
have found that $\Gamma(z=10) = 0.005 - 0.007$ at 1 GeV$^2$. 

As far as moments of twist-2 parton distributions are concerned, lattice
calculations have produced already quite satisfactory predictions for the first
two moments of unpolarized and polarized $C = + 1$ and $C = - 1$ quark
distributions \cite{Latt96}, but extensions to matrix elements of gluonic
operators are not yet as reliable \cite{Glue96}.  Hence, to illustrate the
above procedure we have computed the second and fourth moment of the polarized
gluon distribution at a scale $\mu^2 \approx 1$ GeV$^2$ using QCD sum rules
\cite{SVZ}, suitably tailored for the calculation of matrix elements of QCD
operators \cite{many}.

For this purpose we consider the three-point correlation function
\begin{equation}
I_{\Delta G}= i^2\!\int d^4 x\, e^{iq\cdot x}\int d^4y \, e^{ip\cdot y}
\langle 0| T[ \eta_G(x) O(y+\Delta/2;y-\Delta/2) \bar{\eta}_G(0) ]
|0\rangle  
\label{thetacor}
\end{equation}
of the operator (\ref{defO}) with isospin 1/2 interpolating currents
$\eta_G(x)$ which contain explicitly gluonic degrees of freedom:
\begin{eqnarray}
\eta_G(x) & = & \frac{2}{3} (\eta_{\rm old}(x) - \eta_{\rm ex}(x)), 
\nonumber \\
\eta_{\rm old}(x) & = & \epsilon^{abc} (u^{a T}(x)C\gamma_\mu u^b(x)) \gamma_5
\gamma^\mu \sigma_{\alpha\beta}
\left[ {\rm g} G^{\alpha\beta}(x) d(x) \right]^c,
\nonumber \\
\eta_{\rm ex}(x) & = & \epsilon^{abc} (u^{a T}(x)C\gamma_\mu d^b(x)) \gamma_5
\gamma^\mu \sigma_{\alpha\beta}\left[ {\rm g} G^{\alpha\beta}(x) u(x) \right]^c
\, . 
\label{etaG}
\end{eqnarray}
The current $\eta_G(x)$ was first introduced in \cite{Braun92} to compute the
first moment of the unpolarized gluon density. Since then it has been applied
to estimate higher twist corrections to the Bjorken sum rule \cite{Stein95},
and to calculate the nucleon matrix element of the pseudo-scalar gluon density
$\frac{\alpha_S}{\pi}G_{\mu\nu}{\tilde G}^{\mu\nu}$ \cite{Belitskii96}, giving
in both cases reasonable results. From a technical point of view the main
advantage in using the current $\eta_G$ is that one avoids UV divergences
which occur when gluons are generated perturbatively and which survive the
Borel transformation.  Although in principle we are interested only in the
first two local operators which contribute to the Taylor expansion of
(\ref{defO}) around $\Delta = 0$, it turns out that one can account for the
non-local operator $O(y+\Delta/2;y-\Delta/2)$ as a whole at no additional
expense.  To avoid significant non-perturbative contributions from large
t-channel distances we have simply kept $Q^2 = - q^2 \sim 1 - 3 $ GeV$^2$ in
the Euclidean domain, and extrapolate numerically at the end to $Q^2 = 0$.  The
kinematics is chosen such that $q \cdot \Delta = 0$ and $q^2 = -
q_\perp^2$, which yields $p \cdot \Delta = (p+q) \cdot \Delta = z$.  

We focus on the contribution to the correlator (\ref{thetacor}) of the form:
\begin{equation}
\gamma_5 {\hat p} (p\cdot n)^2 
\int_0^1 du \,T(p^2,(p+q)^2,Q^2, u) \sin(u z),
\end{equation}
where ${\hat p} = p_\mu \gamma^\mu$.  The invariant function
$T(p^2,(p+q)^2,Q^2, u)$ can be projected out uniquely by taking the 
trace of $I_{\Delta G}$ with $\frac{1}{4} \hat n \gamma_5$.  
Identifying the contribution from nucleon
intermediate states to the three-point correlation function (\ref{thetacor})
leads to:
\begin{eqnarray}
\frac{1}{4}  
\frac{\mbox{Tr} \left( \hat n \gamma_5 I_{\Delta G} \right)}
{ (p\cdot n)^3} &=&
\int_0^1 du \,T(p^2,(p+q)^2,Q^2, u) \sin(u z),\nonumber \\
&=&  
\frac{\lambda_G^2 \,m_N^4}{(m_N^2 - p^2)(m_N^2 - (p+q)^2)} 
{\tilde \Gamma}(z,Q^2) + \dots,
\label{nucleon}
\end{eqnarray}
where the form-factor ${\tilde \Gamma}(z;Q^2)$ coincides at $Q^2 = 0$ with the
Ioffe-time distribution $\Gamma(z)$ in eq.(\ref{DelG-Ioffe}).  In
(\ref{nucleon}) the nucleon polarization has been fixed parallel to the $\hat
3$-axis such that $s \cdot \Delta = p \cdot \Delta$, and ellipses denote
contributions from higher resonances and the continuum.  Note that here and in
the following we have suppressed the explicit reference to the renormalization
scale $\mu^2$ which is understood to be approximately equal to $1$ GeV$^2$.
The overlap of the interpolating current $\eta_G$ with the nucleon state,
\begin{equation}
\langle 0|\eta_G|p,s\rangle = m_N^2 \lambda_G u(p,s) \, ,
\label{overlap}
\end{equation} 
at the scale $\mu^2 \sim 1$ GeV$^2$ has been determined in Ref.\cite{Braun92}.

The invariant function $T(p^2,(p+q)^2,Q^2,u)$ admits a double spectral
representation \cite{many}:
\begin{equation}
T(p^2,(p+q)^2,Q^2,u) = 
\int \frac{ds_1}{s_1-p^2}
\int \frac{ds_2}{s_2-(p+q)^2} \;\rho(s_1,s_2,Q^2,u).
\label{disp}
\end{equation}
Therefore, after expanding the left- and right-hand side of eq.(\ref{nucleon}) 
around $\Delta = 0$ we obtain:
\begin{eqnarray} \label{nucleon1}
T_l(p^2,(p+q)^2,Q^2) & = & \int \frac{ds_1}{s_1-p^2}
\int \frac{ds_2}{s_2-(p+q)^2} \;\rho_l(s_1,s_2,Q^2) \nonumber \\
& = & \frac{\lambda_G^2 \,m_N^4}{(m_N^2 - p^2)(m_N^2 - (p+q)^2)} 
\Gamma_l(Q^2) + \dots,\quad l =2,4 \dots . 
\end{eqnarray}
Here $\rho_l(s_1,s_2,Q^2) = \int_0^1 du \,u^{l-1}\rho(s_1,s_2,Q^2,u)$ is the
$l$-th moment of the spectral density.  A comparison with eq.(\ref{twist2})
shows that in the limit $Q^2 \rightarrow 0$ the moments $\Gamma_l(Q^2)$ are
identical to the moments of the polarized gluon distribution.  To eliminate the
contributions from higher resonances and the continuum we limit, according to
the standard procedure, the integrals over $s_1$ and $s_2$ by the continuum
threshold $s_0$ and subsequently perform a Borel transformation in both $p^2$
and $(p+q)^2$, arriving at the sum rule:
\begin{equation}  \label{moments_Borel}
\Gamma_l(Q^2) = 
\frac{e^{m_N^2/M^2}}{\lambda_G^2 m_N^4} 
\int_0^{s_0} d s_1 \int_0^{s_0} d s_2 \,e^{-(s_1+s_2)/2M^2} 
\rho_l(s_1,s_2,Q^2) \, ,
\end{equation}
Note that the continuum threshold $s_0$ and the Borel parameter $M^2$ should in
principle be kept at their values fixed by the two-point nucleon sum rule
\cite{Braun92}.

In the following we concentrate on the sum rules for the first two moments of
$\Delta G(u)$.  In the calculation of the spectral function
$\rho(s_1,s_2,Q^2,u)$ we have restricted ourselves, as in \cite{Braun92}, to
the perturbative and the four-quark condensate contributions only. Explicitly,
we obtain:
\begin{eqnarray}
\rho_4^{\rm pert}(s_1,s_2,Q^2) &=& 
\frac{7 \alpha_s}{2880 \pi^5} \frac{s_1^4 s_2^4 Q^{10}}{R^{13/2}} 
\left(5 (s_1 + s_2) R - 22 s_1 s_2 Q^2\right), 
\\
\rho_2^{\langle \bar q q \rangle}(s_1,s_2,Q^2) &=&  
\frac{280\alpha_s}{27\pi} \,\langle \bar q q \rangle^2\, 
\frac{s_1 s_2 \Delta_q Q^8}{R^{9/2}} 
\left(R + 7 s_1 s_2 \right), 
\\
\rho_4^{\langle \bar q q\rangle}(s_1,s_2,Q^2)  &=&  
\frac{392\alpha_s}{27\pi} \,\langle \bar q q \rangle^2 \, 
\frac{s_1 s_2 \Delta_q Q^{12}}{R^{13/2}} 
\left(R^2  + 18 R  s_1 s_2 + 66 s_1^2 s_2^2\right) \, , 
\end{eqnarray}
where $\Delta_q = s_1 + s_2 + Q^2$ and $R = \Delta_q^2 - 4 s_1 s_2$. The
expression for $\rho_2^{\rm pert}(s_1,s_2,Q^2)$ is unfortunately too long to be
quoted here. It is interesting to note that in this \linebreak approximation
the calculation of the three-point function with the operator \linebreak $n_\mu
n_\nu \mbox{Tr}\,
G^{\mu\xi}(\Delta/2)[\Delta/2;-\Delta/2]G_\xi^\nu(-\Delta/2)$, determining the
Ioffe-time distribution of unpolarized gluons, results in exactly the same
spectral density $\rho(s_1,s_2,Q^2,u)$. In numerical calculations we have used
$-(2\pi)^2\langle \bar q q\rangle \simeq 0.67\,\mbox{GeV}^3$ which is the
standard value of the quark condensate at a scale of $1 \, \mbox{GeV}^2$.

To determine $\Gamma_l\equiv \Gamma_l(Q^2=0)$ we have fitted 
$\Gamma_2(Q^2)$ and
$\Gamma_4(Q^2)$ in the interval $1 \le Q^2 \le 3$ GeV$^2$ with the ansatz:
\begin{equation}
\Gamma_l(Q^2) =  \frac{\Gamma_l}{[1+Q^2/\mu^2]^3}\, ,
\label{fit}
\end{equation}
as suggested by  spectator counting rules. To find the dependence of our
results on $s_0$ and $M^2$ we have varied them around the standard values known
from the two-point nucleon sum rule. The estimates for $\Gamma_2$ and
$\Gamma_4$ obtained in this way are collected in Table 1. We observe a moderate
dependence on the sum rule parameters.  Taking $\sqrt{s_0} = 1.5$ GeV and $M^2
= 1$ GeV$^2$ we get $\Gamma_2 = 0.34$ and $\Gamma_4 = 0.24$ at the scale
$\mu^2 \sim 1$ GeV$^2$.

\begin{table}[t]
{\hfill
\begin{tabular}{|c|c|c|c|} \hline
 $s_0^{1/2}$ [GeV] &  $M^2$ [GeV$^2$]& $\Gamma_2$ & $\Gamma_4$  \\ \hline
 1.4         &  1     & 0.289      & 0.205      \\ \hline
 1.5         &  1     & 0.340      & 0.240      \\ \hline
 1.6         &  1     & 0.387      & 0.265      \\ \hline
 1.4         &  2     & 0.312      & 0.217      \\ \hline
 1.5         &  2     & 0.389      & 0.273      \\ \hline
 1.6         &  2     & 0.477      & 0.335      \\ \hline
\end{tabular}
\hfill}
\caption{$\Gamma_2$ and $\Gamma_4$ for various values of sum rule parameters.}
\label{table1}
\end{table}

With the above results at hand, we obtain the Ioffe-time distribution
$\Gamma(z,\mu^2 \sim 1\mbox{GeV}^2)$ in the small $z$ domain 
by the expansion in eq.(\ref{eq:Ioffe_approx}).  
Together with our estimate for $\Gamma(z=10)$
discussed above, we get $\Delta G \approx 2.0 \pm 0.5$ 
at the scale $\mu^2 \sim 1$ GeV$^2$.  
The error estimate results from the quality of the sum rule. 
This value is not far from $ \Delta G(1\,GeV^2) = 1.7$ for the
parameterization of Chiappetta et al. \cite{Chiap}, and it supports the recent
analysis \cite{Forte96} of the $Q^2$-dependence of polarized structure function
data which yields $\Delta G(1\,GeV^2) = 1.5 \pm 0.8$.  Contributions to
$\Gamma(z)$ from $z \gsim 10$ can change the value of $\Delta G$ by about 10\%.
On the other hand the present calculation overestimates the true
values of $\Gamma_l$. This can be seen for example from our prediction for
$\Gamma_4$ being larger then the experimentally determined
value of the third moment of the unpolarized gluon distribution 
\cite{Ross96}, which contradicts unitarity.  
Our result for $\Delta G$ should therefore be interpreted as 
an upper limit.  

To summarize, we have found that for sufficiently smooth polarized gluon
distributions the gluon polarization in the nucleon can be estimated to good
accuracy from the first two even moments of the polarized gluon density,
together with its behavior at large Ioffe-times, or equivalently at small
values of Bjorken variable $x$. We have calculated the former within the
framework of QCD sum rules. Taking into account the perturbative and the
four-quark condensate contribution we found the corresponding spectral function
to be similar to the spectral function for the gluonic operator determining the
unpolarized gluon distribution.  Consequently, since the contribution of gluons
to the nucleon momentum is around $50\%$ even at low scales $\mu^2 \sim
1\,GeV^2$, also the polarized gluon distribution is expected to be large.
Combining the calculated moments with the color coherence of gluon couplings we
have obtained $\Delta G(\mu^2 \sim 1\,GeV^2) \approx 2$, which is
acceptable from a phenomenological point of view. Finally we want to
emphasize the need for a calculation of the first two moments of the polarized
gluon density on the lattice.  Unfortunately, even an order of magnitude
estimate of $\Gamma_4$ requires the treatment of a
symmetric traceless tensor of rank 5 and therefore appears to be quite
difficult.

\vskip 1 cm

{\bf Acknowledgments} We want to acknowledge numerous discussions
with V. Braun about calculations presented in this paper. This work was
supported in part by BMBF, KBN grant 2~P03B~065~10, and German-Polish exchange
program X081.91.

\vfill
\eject

\clearpage
\begin{minipage}{16cm}
\bild{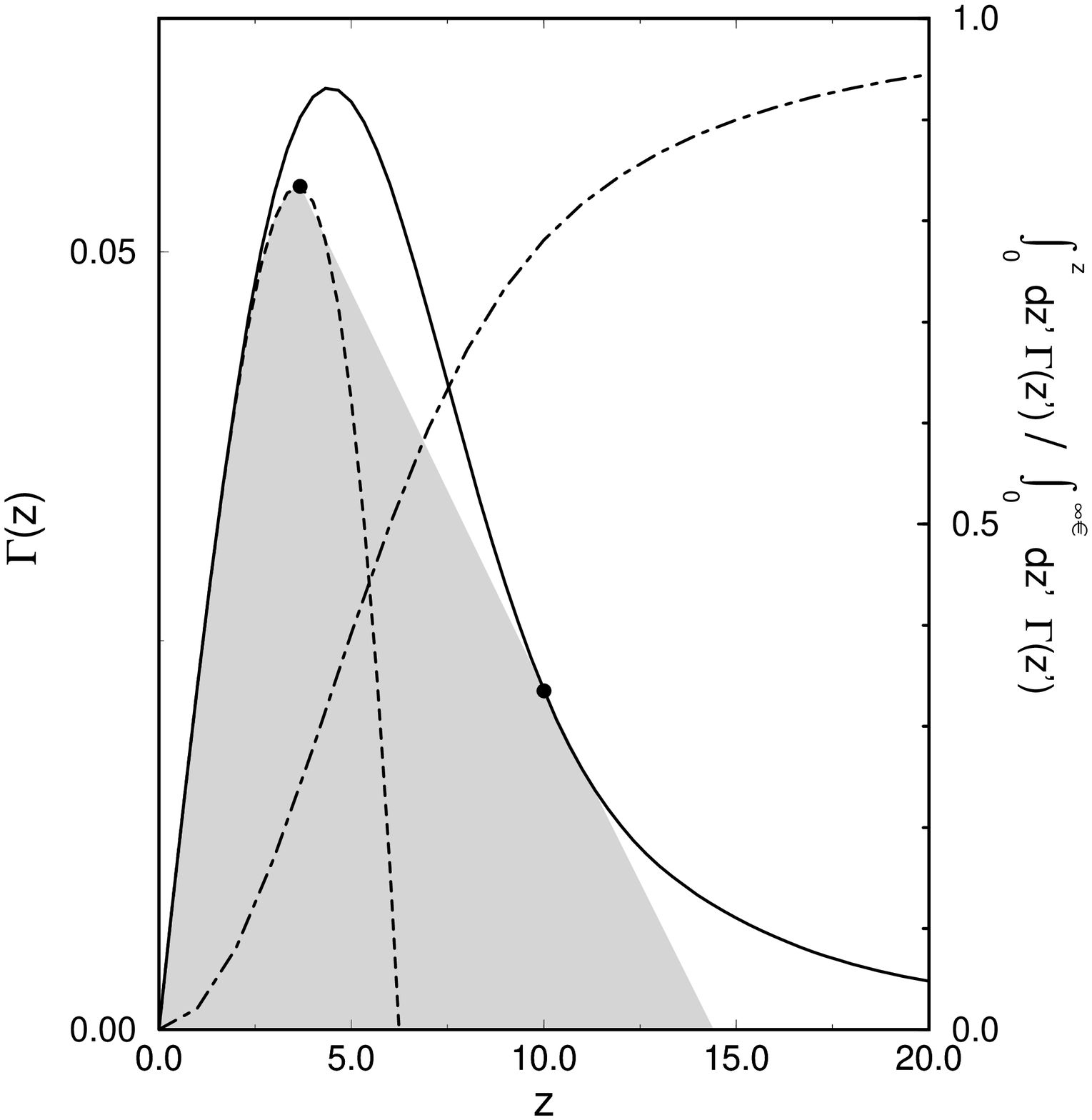}{13}
\end{minipage}
\begin{description}
  
\item[Fig.~1] The solid line represents the Ioffe-time distribution $\Gamma(z)$
  of the polarized gluon density of Brodsky et al. \cite{Bro95}.  The
  approximate Ioffe-time distribution defined by the moments $\Gamma_2$ and
  $\Gamma_4$ (\ref{eq:Ioffe_approx}) is shown by the dotted line.  The fraction
  of the first moment integral (\ref{defDG}) as a function of the upper
  integration limit $z$ is given by the dashed-dotted line.  The shaded area
  yields the approximation to $\Delta G$ as explained in the text.
\end{description}

\begin{minipage}{16cm}
\bild{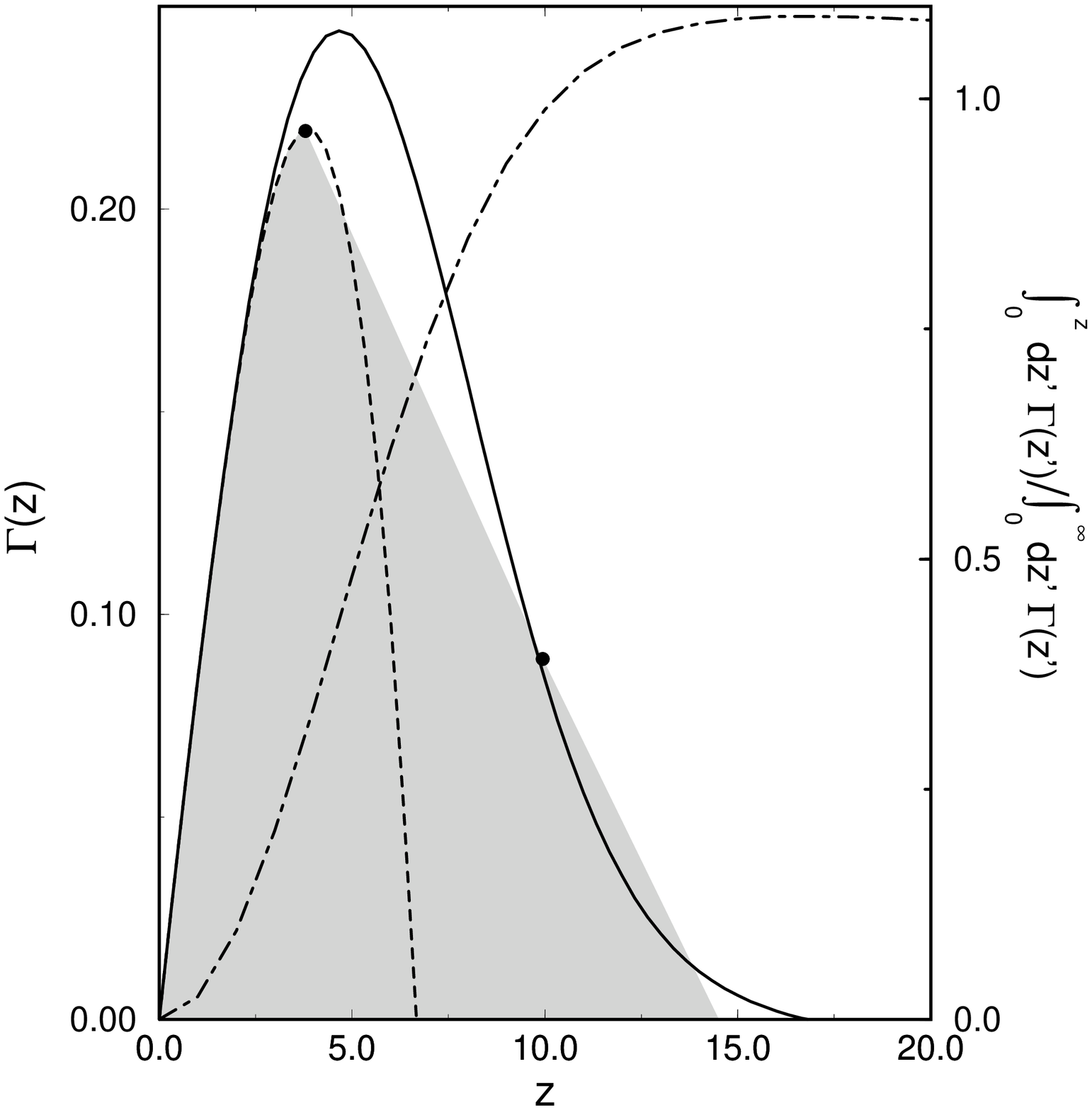}{13}
\end{minipage}
\begin{description}
  
\item[Fig.~2] 
  As  Figure 1 for the polarized gluon distribution of 
  Chiappetta et al. \cite{Chiap}.

\end{description}

\end{document}